\documentclass[aps,prl,reprint,preprintnumbers,superscriptaddress,nofootinbib]{revtex4-2}
\usepackage{graphicx}
\usepackage{dcolumn}
\usepackage{bm}
\usepackage[pdfencoding=auto, psdextra]{hyperref}
\usepackage{amsmath}

\usepackage{epstopdf}
\usepackage{epsfig}
\usepackage{graphics}
\usepackage{xspace}
\usepackage{slashed}
\usepackage{xcolor}
\usepackage{longtable}
\usepackage[small]{subfigure}
\usepackage[normalem]{ulem}
\usepackage{multirow}
\usepackage[normalem]{ulem}

\newcommand{\beq}{\begin{equation}}
\newcommand{\eeq}{\end{equation}}
\newcommand{\units}{\times 10^{-10}}

\begin{document}

\title{\boldmath Data-driven determination of the light-quark connected component of the intermediate-window contribution to the muon \texorpdfstring{$g-2$}{g-2}}

\author{Genessa Benton}
\affiliation{Department of Physics and Astronomy, San Francisco State University,\\
San Francisco, CA 94132, USA}
\author{Diogo Boito}
\affiliation{Instituto de F\'isica de S\~ao Carlos, Universidade de
S\~ao Paulo, CP 369, 13560-970, S\~ao Carlos, SP, Brazil}
\author{Maarten Golterman}
\affiliation{Department of Physics and Astronomy, San Francisco State University,\\
San Francisco, CA 94132, USA}

\author{Alexander Keshavarzi}
\affiliation{Department of Physics and Astronomy, The University of Manchester,\\
Manchester M13 9PL, United Kingdom}

\author{Kim Maltman}
\affiliation{Department of Mathematics and Statistics,
York University,\\ Toronto, ON Canada M3J~1P3}
\affiliation{CSSM, University of Adelaide, Adelaide, SA~5005 Australia}

\author{Santiago Peris}
\affiliation{Department of Physics and Astronomy, San Francisco State University,\\
San Francisco, CA 94132, USA}
\affiliation{Department of Physics and IFAE-BIST, Universitat Aut\`onoma de Barcelona,\\
E-08193 Bellaterra, Barcelona, Spain}

\date{\today}

\begin{abstract}

We present the first data-driven result for $a_\mu^{\rm win,lqc}$, the
isospin-limit light-quark connected component of the intermediate-window
Hadronic-Vacuum-Polarization contribution to the muon anomalous
magnetic moment. Our result, $(198.8\pm 1.1)\times 10^{-10}$, is in
significant tension with eight recent mutually compatible high-precision
lattice-QCD determinations, and provides enhanced evidence for a puzzling
discrepancy between lattice and data-driven determinations of the
intermediate window quantity, one driven largely by a difference in the
light-quark connected component.

\end{abstract}

\maketitle

\vspace{0.5cm}
Since the pioneering work of Schwinger~\cite{Schwinger:1948iu}, and the
subsequent experimental confirmation of his result~\cite{Kusch:1948mvb},
the study of the lepton magnetic moments, $\vec \mu$, has played a
central role in the development of Quantum Electrodynamics (QED) and,
later, in that of the Standard Model (SM) of particle physics.
The anomalous magnetic moment of the muon, $a_\mu=(g_\mu-2)/2$, defined
in terms of the muon charge, $q$, mass, $m_\mu$, and spin, $\vec s$, as
\beq
\vec \mu =g_\mu \left( \frac{q}{2m_\mu} \right)\vec s
\eeq
is, at present, one of the most accurately known quantities in
physics. The new 2021 and 2023 Fermilab E989 measurements of $a_\mu$,
performed by the Muon $g-2$ collaboration, reached a precision of
0.20 ppm~\cite{Muong-2:2021vma,Muong-2:2021ojo,Muong-2:2023cdq}, and are fully compatible
with the previous BNL E821 experiment results~\cite{Muong-2:2006rrc}. As
a consequence, the experimental average of $a_\mu$ is known nowadays to
0.19 ppm. In order to stringently test the SM, the theory prediction for
this quantity must reach a similar level of accuracy.

In 2020, the $g-2$ Theory Initiative released a white paper
(WP)~\cite{Aoyama:2020ynm} where, based on the works of
Refs.~\cite{Aoyama:2012wk,Aoyama:2019ryr,Czarnecki:2002nt,Gnendiger:2013pva,Davier:2017zfy,Keshavarzi:2018mgv,Davier:2019can,Keshavarzi:2019abf,Colangelo:2018mtw,Hoferichter:2019mqg,Hoid:2020xjs,Kurz:2014wya,Melnikov:2003xd,Masjuan:2017tvw,Colangelo:2017qdm,Colangelo:2017fiz,Hoferichter:2018dmo,Hoferichter:2018kwz,Gerardin:2019vio,Bijnens:2019ghy,Colangelo:2019lpu,Colangelo:2019uex,Colangelo:2014qya,Blum:2019ugy},
the SM expectation for $a_\mu$ was determined to 0.37 ppm. The SM result
is a sum of pure QED, electroweak and Higgs physics, hadronic vacuum
polarization (HVP), and hadronic light-by-light scattering (HLbL)
contributions. The latter two, which involve Quantum Chromodynamics (QCD),
are particularly difficult to assess. The final uncertainty in the WP
SM result for $a_\mu$ is strongly dominated by the HVP contribution,
$a_\mu^{\rm HVP}$. It is thus a fundamental task to control this
contribution as well as possible.

The $a_\mu$ assessment of the WP is based on data-driven evaluations of
the HVP contribution~\cite{Davier:2017zfy,Davier:2019can,Keshavarzi:2018mgv,Keshavarzi:2019abf}.
As is well known, this result showed a $4.2\sigma$ tension with the
then-current experimental average, motivating many works aimed at
finding potential beyond-the-SM (BSM) explanations for this
disagreement. Many developments, however, have taken place since the
publication of the the
WP assessment. In particular, the
Budapest-Marseille-Wupertal (BMW) collaboration has published a
complete, sub-percent lattice-QCD evaluation of $a_\mu^{\rm HVP}$~\cite{Borsanyi:2020mff}.
If this result is used in the SM assessment of $a_\mu$, the outcome is
compatible with the experimental average at the 1.5$\sigma$ level. In
this situation, in order to conclude whether the discrepancy between
the SM assessment(s) of $a_\mu$ and experimental results is due to BSM
effects, it is, first of all, crucial to understand the discrepancy
between data-driven and lattice-QCD based HVP results.

The complete computation of the HVP in lattice QCD is a very challenging
task and, as of today, only the BMW result is sufficiently precise to
allow for a detailed comparison with data-driven HVP determinations. This
has motivated the introduction of the window quantities by the RBC/UKQCD
collaboration \cite{RBC:2018dos}. The so-called intermediate window,
which cuts out short- and long-distance regions of the $a_\mu^{\rm HVP}$
integral on the lattice-QCD side, is of particular importance because it
significantly suppresses the associated systematic lattice uncertainties
related to continuum limit extrapolation and finite volume effects.
The intermediate-window contribution to $a_\mu^{\rm HVP}$, henceforth
$a_\mu^{\rm win}$, is now known, with excellent precision and very
good agreement between results from different groups, from
four different lattice-QCD collaborations~\cite{Borsanyi:2020mff,Ce:2022kxy,ExtendedTwistedMass:2022jpw,Blum:2023qou}.
A recent data-driven assessment of $a_\mu^{\rm win}$ using electroproduction input~\cite{Colangelo:2022vok},
however, shows an important tension with these lattice results
(see e.g.~Ref.~\cite{Wittig:2023pcl}). (This
tension is reduced if, instead, one uses $\tau$ data for the two-pion
contribution~\cite{Masjuan:2023qsp}, estimating the required model-dependent isospin-breaking
corrections with the models discussed in Ref.~\cite{Miranda:2020wdg}.)

\begin{figure}[!t]
\includegraphics[width=1.0\columnwidth]{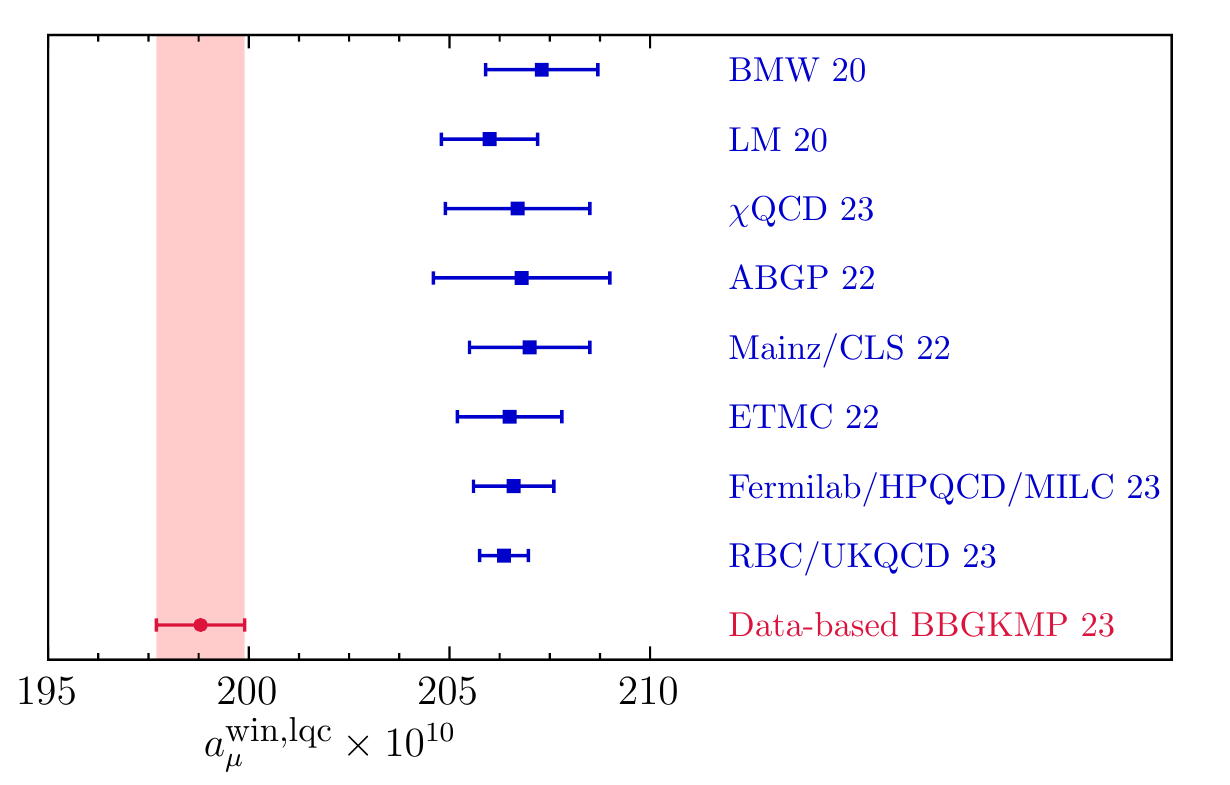}
\caption{Comparison of our final result (BBGKMP 23), Eq.~(\ref{eq:final-res}),
with lattice results for $a_\mu^{\rm win,lqc}$ from \cite{Borsanyi:2020mff}
(BMW 20), \cite{Lehner:2020crt} (LM 20), \cite{Wang:2022lkq} ($\chi$QCD 23),
\cite{Aubin:2022hgm} (ABGP 22), \cite{Ce:2022kxy} (Mainz/CLS 22),
\cite{ExtendedTwistedMass:2022jpw} (ETMC 22), \cite{FermilabLatticeHPQCD:2023jof} (FHM 23),
and \cite{Blum:2023qou} (RBC/UKQCD 23)}
\label{fig:results}
\end{figure}

The lattice evaluation of $a_\mu^{\rm HVP}$ is split into several building
blocks with the dominant contributions arising from the isospin-limit
(defined by taking $m_\pi=m_{\pi^0}$) light- and strange-quark connected
and disconnected parts, with additional, smaller, contributions from
charm and bottom quarks.
Isospin breaking (IB) effects, both of electromagnetic (EM) and strong (SIB)
origin, are accounted for perturbatively, keeping terms to first
order in an expansion in the fine-structure constant $\alpha$ and the
up-down quark-mass difference $m_u-m_d$. The light-quark connected (lqc)
contribution to $a_\mu^{\rm win}$ in the isospin symmetric limit,
denoted $a_\mu^{{\rm win, lqc}}$, is known now with very good
precision from eight different lattice determinations (see the blue data
points in Fig.~\ref{fig:results}). These eight determinations are all
in excellent agreement and have small relative errors, ranging from 0.3
to 1.1\%. Since this contribution gives about 87\%
of $a_\mu^{\rm win}$, and appears to be under good control, given the
agreement among the eight different lattice determinations, it is
highly desirable to obtain a precise data-driven estimate of
$a_\mu^{{\rm win,lqc}}$ in order to further scrutinize the
discrepancy between lattice-QCD and data-driven determinations of
$a_\mu^{\rm HVP}$. It is the aim of this letter to present this estimate.

We turn now to a short review of the theoretical framework for our
data-driven determination of $a_\mu^{{\rm win, lqc}}$. To be able to
compute $a_\mu^{\rm HVP}$ on the lattice-QCD side, one determines the
Euclidean-time zero-momentum two-point correlation function given by
\begin{eqnarray}
C(t)&=&\frac{1}{3}\sum_{i=1}^3\int d^3x \langle j_i^{\rm EM}(\vec{x},t)
j_i^{\rm EM}(0)\rangle
\nonumber
\\&=&\frac{1}{2}\int_{m_\pi^2}^\infty ds\,\sqrt{s}\,
e^{-\sqrt{s}t}\,\rho_{\rm EM}(s)\quad (t>0),
\end{eqnarray}
where $m_\pi$ is the neutral pion mass, $j_\mu^{\rm EM}(x)$ is
the EM current, and $\rho_{\rm EM}$ is the associated inclusive
hadronic spectral function. In terms of $C(t)$, the intermediate-window
contribution to $a_\mu^{\rm HVP}$ is given by
\beq\label{eq:amuwin-latt}
a_\mu^{\rm win}=2\int_0^\infty dt\,w(t) W_{\rm win}(t) C(t) ,
\eeq
where the function $w(t)$ can be obtained from its counterpart in
$s$-space~\cite{Bernecker:2011gh} and $W_{\rm win}(t)$ is the weight function
associated with the RBC/UKQCD intermediate-window~\cite{RBC:2018dos},
defined as
\beq
W_{\rm win}(t)=\frac{1}{2}\left(\tanh\frac{t-t_0}{\Delta}-\tanh
\frac{t-t_1}{\Delta}\right),
\eeq
with $t_0=0.4$~fm, $t_1=1.0$~fm, and $\Delta = 0.15$~fm. The
corresponding expression for $a_\mu^{\rm HVP}$ is obtained by removing
the factor $W_{\rm win}(t)$ from Eq.~(\ref{eq:amuwin-latt}).
Due to the presence of $W_{\rm win}(t)$ in Eq.~(\ref{eq:amuwin-latt})
the short- and long-distance contributions to the integral are strongly
suppressed.

Since we are concerned with the data-driven determination of
$a_\mu^{{\rm win,lqc}}$, we need the data-driven (or dispersive)
counterpart to Eq.~(\ref{eq:amuwin-latt}) which is
\beq\label{eq:amuwin-disp}
a_\mu^{\rm win} = \frac{4\alpha^2m_\mu^2}{3}\int_{m_\pi^2}^\infty ds\,
\frac{\hat{K}(s)}{s^2} \,\widetilde{W}_{\rm win}(s)\,\rho_{\rm EM}(s),
\eeq
where $\hat K(s)$ is a well known, slowly varying, kernel function~\cite{Brodsky:1967sr,Lautrup:1968tdb}
(see Ref.~\cite{Aoyama:2020ynm} for the explicit expression) and
$\widetilde W_{\rm win}(s)$ is the $s$-space representation of
$W_{\rm win}(t)$,
\beq
\widetilde W_{\rm win} (s)=\frac{\int_0^\infty dt\,W_{\rm win}(t)\,
w(t)\,e^{-\sqrt{s}t}}{\int_0^\infty dt\,w(t)\,e^{-\sqrt{s}t}}.
\eeq
Here, when evaluating the different contributions to
Eq.~(\ref{eq:amuwin-disp}) in the exclusive-mode region, we
employ the data compilation of Refs.~\cite{Keshavarzi:2018mgv,Keshavarzi:2019abf} (KNT19
in what follows).

Our goal is to isolate the lqc contribution to the full result
of Eq.~(\ref{eq:amuwin-disp}). This can be achieved employing an idea
first implemented with sufficient precision in
Refs.~\cite{Boito:2022dry,Boito:2022rkw}, where the foundations of our
method are laid out in detail. We start from the usual decomposition
of the three-flavor EM current into its $I=1$ and $I=0$ parts, which
produces analogous decompositions of $C(t)$ and $\rho_{\rm EM}$
into $I=1$, $I=0$ and mixed-isospin (MI) parts. In the isospin
limit, the contribution associated with the $I=1$ current contains
only light-quark connected contributions and one has
\beq\label{eq:lqc-part}
\rho_{\rm EM}^{\rm lqc} = \frac{10}{9} \rho_{\rm EM}^{I=1}(s).
\eeq
The data-driven estimate of $a_\mu^{{\rm win,lqc}}$ thus requires
the identification of the $I=1$ component of $\rho_{\rm EM}$. This can be
accomplished, assuming isospin-symmetry and using KNT19 data, on a
channel-by-channel basis, in the KNT19 exclusive-mode region,
$\sqrt{s}\le 1.937$ GeV.

There are two classes of such contributions.
The first, and dominant one, consists of contributions from modes
with well-defined, positive $G$ parity. Such modes have $I=1$ and
thus contribute to the lqc component in the isospin limit. Contributions
from such ``unambiguous'' modes constitute the main ingredient in our
determination of $a_\mu^{{\rm win,lqc}}$. The remaining contributions
come from higher-threshold modes with no well-defined $G$ parity. For
these ``ambiguous'' modes, and especially for the dominant such
channels --- $K\bar K$ and $K\bar K \pi$ --- one resorts to external
information, whenever available, in order to identify, as accurately
as possible, the $I=1$ component of the experimental $I=0+1$ sum.
The third ingredient is perturbative QCD supplemented with an estimate of
Duality Violation contributions, which we use in the inclusive
region, $\sqrt{s}> 1.937$~GeV. Finally, to the sum of the results
thus obtained one must apply isospin-breaking (IB) corrections since the
experimental data, inevitably, contain IB contributions. Only after
these corrections have been applied can the data-based result be compared
directly with isospin-limit lattice $a_\mu^{{\rm win,lqc}}$ results.
We now detail how we treat each of the four aforementioned contributions.

We start with the unambiguous modes, which give the dominant contribution
to $a_\mu^{{\rm win,lqc}}$. The results are obtained with
Eq.~(\ref{eq:amuwin-disp}) using the KNT19 spectra for the different
exclusive-mode contributions to $\rho_{\rm EM}(s)$. There are 13
such $I=1$ channels, with the largest contribution, by far, arising
from the $\pi^+\pi^-$ channel, which contributes
$144.15(49)\times 10^{-10}$ to $[a_\mu^{\rm win}]_{I=1}$.
The results for the different modes are given in
Table~\ref{tab:unambiguous-modes}. The sum over all
$G$-parity-unambiguous modes gives a total $I=1$ contribution
to $a_\mu^{{\rm win}}$ of $168.24(72)\units$. From
Eq.~(\ref{eq:lqc-part}), the final contribution of all unambiguous
channels to $a_\mu^{\rm win,lqc}$ is then
\begin{eqnarray}
\label{eq:G-par-modes}
[a_\mu^{\rm win,lqc}]_{G-{\rm par}}
&=&186.93(80)\units.
\end{eqnarray}

\begin{table}
\caption{\label{tab:unambiguous-modes}
Contributions from $G$-parity unambiguous modes to $a_\mu^{\rm win}$ for
$\sqrt{s}\leq 1.937$~GeV obtained from KNT19~\cite{Keshavarzi:2019abf}
exclusive-mode spectra. All entries in units of $10^{-10}$. }
\begin{ruledtabular}
\begin{tabular}{ll}
$I=1$ modes $X$&$[a_\mu^{\rm win}]_X\times 10^{10}$ \\
\hline
low-$s$ $\pi^+ \pi^-$& 0.02(00)\quad\\
$\pi^+ \pi^-$& 144.13(49)\\
$2\pi^+ 2\pi^-$& 9.29(13)\\
$\pi^+ \pi^- 2\pi^0$& 11.94(48)\\
$3\pi^+ 3\pi^-$ (no $\omega$)& 0.14(01)\\
$2\pi^+2\pi^-2\pi^0$ (no $\eta$)& 0.83(11)\\
$\pi^+\pi^- 4\pi^0$ (no $\eta$)& 0.13(13)\\
$\eta \pi^+ \pi^-$& 0.85(03)\\
$\eta 2\pi^+ 2\pi^-$& 0.05(01)\\
$\eta \pi^+\pi^- 2\pi^0$& 0.07(01)\\
$\omega (\rightarrow \pi^0\gamma)\pi^0$& 0.53(01)\\
$\omega (\rightarrow {\rm npp})3\pi$& 0.10(02)\\
$\omega \eta \pi^0$& 0.15(03)\\
\hline
TOTAL& 168.24(72)\\
\end{tabular}
\end{ruledtabular}
\end{table}

We turn next to our treatment of ambiguous-mode contributions,
which follows the general strategy outlined in Sec. IV of
Ref.~\cite{Boito:2022rkw}. For some of these contributions, notably
those of the numerically dominant $K\bar K$ and $K\bar K\pi$ channels,
external experimental information can be used in separating the desired
$I=1$ component from the experimental $I=0+1$ sum. Modes for which
external experimental information is not available have much smaller
contributions. For these modes, one employs a maximally conservative
$I=1/0$ separation, based on the observation that the $I=1$ part of
the mode-$X$ contribution to $\rho_{\rm EM}(s)$ must lie between 0 and
the full $I=0+1$ contribution obtained from the KNT19 spectrum for that
mode. The contribution of ambiguous mode $X$ to
$a_\mu^{\rm win,lqc}$ lies, therefore, in the following range
\beq
\label{eq:max-conser-sep}
[a_\mu^{\rm win,lqc}]_X=\frac{10}{9}\left( \frac{1}{2}\pm \frac{1}{2} \right)
[a_\mu^{\rm win}]_X
= \left( \frac{5}{9}\pm \frac{5}{9} \right)
[a_\mu^{\rm win}]_X.
\eeq

Let us discuss in some detail the significant ambiguous-mode contribution
arising from the $K\bar K$ channels, $K^+K^-$ and $K^0\bar K^0$. Independent
experimental information on the $K\bar K$ contribution to the purely
$I=1$ spectral function can be obtained from the BaBar spectrum for
the decay $\tau\to K^-K^0\nu_\tau$~\cite{BaBar:2018qry}. Using the
conserved vector current (CVC) relation, these results can be used to
determine the $I=1$ $K\bar K$ contribution to $\rho_{\rm EM}(s)$
up to $s=2.7556$~GeV$^2$, and hence the associated contribution
to $\left[a_\mu^{\rm win,lqc}\right]_{K\bar{K}}$, which, using
Eq.~(\ref{eq:lqc-part}), is found to be $10/9\times0.465(29)\units$. For
$s>2.7556$~GeV$^2$, the $I=1$ part is found using KNT19 data and
the maximally conservative treatment of Eq.~(\ref{eq:max-conser-sep}). We
find, for $s>2.7556$~GeV$^2$, a contribution of $10/9\times0.055(55)\units$
to $\left[a_\mu^{\rm win,lqc}\right]_{K\bar{K}}$. From these results one
obtains, for the full exclusive-region $K\bar{K}$ contribution,
\beq
\left[a_\mu^{\rm win,lqc}\right]_{K\bar{K}}=0.578(69)\times 10^{-10}.
\eeq
A similar treatment of the $K\bar K\pi$ modes is possible thanks to the
Dalitz plot analysis of BaBar, which provides a separation of the
$I=1$ and $I=0$ contributions to the $K\bar{K}\pi$ cross
sections\cite{BaBar:2007ceh}. Integrating the BaBar $I=1$ result,
we find
\beq
\left[a_\mu^{\rm win,lqc}\right]_{K\bar{K}\pi}
=0.521(86)\times 10^{-10}.
\eeq
For the $K\bar K 2\pi$ modes, only a small improvement is possible over
the maximally conservative treatment. This is obtained by first subtracting
the small $I=0$ $\phi [\rightarrow K\bar{K}] \pi\pi$ contribution implied
by BaBar $e^+e^-\rightarrow \phi\pi\pi$ cross sections~\cite{BaBar:2011btv},
and applying the maximally conservative treatment only to the residual
$I=0+1$ sum. This leads to the result
\beq
\left[a_\mu^{\rm win,lqc}\right]_{K\bar{K}2\pi} = 0.60(60)\units.
\eeq
The very small (often completely negligible) contributions of the
remaining ambiguous modes ($K\bar{K}3\pi$,
$\omega(\rightarrow {\rm npp})K\bar{K}$,
$\eta (\rightarrow {\rm npp}) K\bar{K}$ (no $\phi$), $p\bar{p}$,
$n\bar{n}$, and low-$s$ $\pi^0 \gamma$ and $\eta \gamma$) (npp =
non-purely pionic) are obtained from the KNT19 spectra using
the maximally conservative separation of Eq.~(\ref{eq:max-conser-sep}).
The total contribution from all $G$-parity-ambiguous exclusive
modes is, finally,
\beq\label{eq:amb-modes}
[a_\mu^{\rm win,lqc}]_{\rm amb.}= 1.74(61)\units,
\eeq
with $1.70(61)\units$ from $K\bar{K}$, $K\bar{K}\pi$ and
$K\bar{K}2\pi$.

In the inclusive region, $\sqrt{s}>1.937$~GeV, we use QCD perturbation
theory, which is known to $\mathcal{O}(\alpha_s^4)$, supplemented
with an estimate for the $\mathcal{O}(\alpha_s^5)$ coefficient, as described
in Refs.~\cite{Boito:2022dry,Boito:2022rkw}. To this result, we add an
estimate of the duality violation (DV) contribution, obtained from
our previous study of the $I=1$ hadronic spectral function in
$\tau\to {\rm hadrons}+\nu_\tau$~\cite{Boito:2020xli}, using the
parametrization of DVs discussed in
Refs.~\cite{Boito:2017cnp,Cata:2005zj,Cata:2008ye}. Since DVs
represent a fundamental limitation of perturbation theory, we use
the resulting central value as the total uncertainty on the
perturbative contribution. This enlarges the uncertainty of
the perturbative contribution without DVs by a factor of about 10,
and should provide a very conservative assessment. The resulting
inclusive-region contribution is then
\beq
\label{eq:pt-QCD-and-DVs}
[a_\mu^{\rm win,lqc}]_{\rm pt. QCD+DVs} = 11.06(16)\units.
\eeq

The fourth and final ingredient in the determination of
$a_\mu^{\rm win,lqc}$ is an evaluation of the EM and SIB
contributions to be subtracted from the data-based results
obtained above before comparison with isospin-symmetric lattice-QCD
results. The general strategy employed for this subtraction
is detailed in Refs.~\cite{Boito:2022dry} and
\cite{Boito:2022rkw}. The main observation is that, to first order in IB,
SIB is present only in the MI component of $\rho_{\rm EM}(s)$.
EM IB, on the other hand, occurs in all of the pure $I=1/0$ and
MI components. The IB correction to $a_\mu^{\rm win,lqc}$
then contains two parts. The first, which appears in the pure
$I=1$ component, is of EM origin. No breakdown of this correction
into individual exclusive-mode contributions is required; an inclusive
determination is sufficient. The situation for the MI contribution is
different since we must estimate the MI contamination on a channel-by-channel
basis, removing from the ``nominally'' $I=1$ results above the
component that arises from $\rho^{\rm MI}_{\rm EM}(s)$. These contributions
are expected to be dominated by the $2\pi$ mode through $\rho-\omega$ mixing
in the process $e^+e^-\to \omega\to \rho\to 2\pi$.

At present, given the absence of complete data-driven estimates for some
potentially important components of the pure $I=1$ EM IB contribution (see,
e.g., the discussion in the Appendix of Ref.~\cite{Boito:2022dry}),
we are forced to rely on the lattice, and employ for this correction
the result obtained by BMW in Ref.~\cite{Borsanyi:2020mff},
\beq
\label{eq:EM-IB}
\Delta_{\rm EM}a_\mu^{\rm win,lqc}=0.035(59)\times 10^{-10}\ .
\eeq
This correction is very small, given the size of other uncertainties,
and we will neglect it in what follows.

The MI contamination to the $2\pi$ exclusive mode was obtained in
Ref.~\cite{Colangelo:2022prz} from $2\pi$ electroproduction data
fitting a dispersive representation of the pion form factor that includes
$\rho-\omega$ mixing. The $2\pi$ MI component is found to be
$\left[ a_\mu^{\rm win}\right]_{\pi\pi}^{\rm MI}=0.83(6)\times 10^{-10}$,
which is about $0.6\%$ of the total $2\pi$ contribution to $a_\mu^{\rm win}$.
Since the MI components of other nominally $I=1$ modes have no analogous narrow-resonance
enhancements, we consider it very safe to assume that their MI total will
not exceed $1\%$ of the sum, $25.68\units$, of their contributions.
To account for the total of the non-$2\pi$-mode MI contaminations
we thus add an uncertainty of $0.26\units$ to the $2\pi$ results of
Ref.~\cite{Colangelo:2022prz}. Using Eq.~(\ref{eq:lqc-part}), this
leads to a MI correction of
\beq
\label{eq:MI-IB-correc}
\Delta^{\rm MI}a_\mu^{\rm win,lqc}
= -0.92(7)_{2\pi}(29)_{{\rm non-}2\pi}\times 10^{-10}.
\eeq

We are now in a position to obtain our final data-driven estimate
for $a_\mu^{\rm win,lqc}$. Adding the contributions from
Eqs.~(\ref{eq:G-par-modes}), (\ref{eq:amb-modes}), (\ref{eq:pt-QCD-and-DVs}),
and applying the IB correction of Eq.~(\ref{eq:MI-IB-correc}), we find, as
our final result,
\beq
\label{eq:final-res}
a_\mu^{\rm win,lqc}=198.8(1.1)\times 10^{-10} .
\eeq
In Fig.~\ref{fig:results}, we compare our data-driven estimate with the
lattice-QCD results from 8 different collaborations. The tension between
the data-driven and lattice results is striking. Assuming, for simplicity,
all errors to be Gaussian, we find tensions ranging from $3.3\sigma$ to
$6.1\sigma$. Our result indicates that the discrepancy between data-driven
and lattice-QCD results for $a_\mu^{\rm win}$ is almost entirely due to
the light-quark connected contribution, which, in turn, is strongly
dominated by the $2\pi$ channel---accounting for about 81\% of the
result of Eq.~(\ref{eq:final-res}). Given this $2\pi$ dominance, it is
relevant to note that recent CMD-3 results for the $e^+e^-\to \pi^+\pi^-$
cross sections~\cite{CMD-3:2023alj}, which are in significant tension with
those of earlier experiments, and known to significantly increase the $2\pi$
contribution to $a_\mu^{\rm HVP}$, would similarly increase our result for
$a_\mu^{\rm win, lqc}$, making it more compatible with lattice determinations.
Since the source of the disagreements between the previously published and new
CMD-3 $2\pi$ results is both presently unclear and the subject of ongoing
study, we refrain from addressing this issue more quantitatively for now.

We note that our final result is based on the KNT19 data compilation. An
equivalent analysis using other dispersive evaluations (e.g.
DHMZ data~\cite{Davier:2017zfy,Davier:2019can})
would be desirable. We remark, however, that for
the lqc contribution to $a_\mu^{\rm HVP}$, which can be obtained based
on publicly available results, KNT19- and DHMZ-based estimates are
in very good agreement~\cite{Boito:2022dry}.

In a companion paper~\cite{Benton:2023fcv} we present results for several other
window quantities, including both the light-quark-connected
and strange-quark-plus-all-disconnected contributions. The latter require the treatment of the $I=0$ sector. The impact of new phenomenological estimates of MI IB corrections in the $3\pi$ channel~\cite{Hoferichter:2023sli} are discussed and we present a comparison between lattice-QCD and phenomenological IB corrections. A preliminary estimate of the potential impact of new CMD-3 results for $e^+e^-\to \pi^+\pi^-$ cross-section~\cite{CMD-3:2023alj} is also given.

\vspace*{0.5cm}

\begin{acknowledgments}
{\bf Acknowledgments:}
We would like to thank Martin Hoferichter and Peter Stoffer for extensive
discussions on isospin breaking.
DB and KM thank San Francisco State University where part of this
work was carried out, for hospitality. This material is based upon work
supported by the U.S. Department of Energy, Office of Science, Office of
Basic Energy Sciences Energy Frontier Research Centers program under Award
Number DE-SC-0013682 (GB and MG). DB's work was supported by the S\~ao Paulo
Research Foundation (FAPESP) Grant No. 2021/06756-6 and by CNPq Grant No.
308979/2021-4.
The work of AK is supported by The Royal Society (URF\textbackslash R1\textbackslash231503), STFC (Consolidated Grant ST/S000925/) and the European Union’s Horizon 2020 research and innovation programme under the Marie Sklodowska-Curie grant agreement No.~858199 (INTENSE).
The work of KM is supported by a grant
from the Natural Sciences and Engineering Research Council of Canada. SP
is supported by the Spanish Ministry of Science, Innovation and Universities
(project PID2020-112965GB-I00/AEI/10.13039/501100011033) and by Departament de
Recerca i Universitats de la Generalitat de Catalunya, Grant No 2021
SGR 00649. IFAE is partially funded by the CERCA program of the Generalitat
de Catalunya.

\end{acknowledgments}

\bibliography{Refs-amuW1lqc}

\end{document}